\documentclass[aps,pra,twocolumn,superscriptaddress,longbibliography]{revtex4-2}

\usepackage{amsmath,amssymb}
\usepackage{graphicx}
\usepackage{bm}
\usepackage[hidelinks]{hyperref}
\usepackage{orcidlink}

\newcommand{\op}[1]{\hat{#1}}

\newcommand{\var}{\mathrm{Var}}
\newcommand{\kO}{k_0}
\newcommand{\Sig}{\Sigma}

\begin{document}

\title{Detection-resolution limits of large-momentum-transfer atom gravimetry}

\author{Asad Ali\,\orcidlink{0000-0001-9243-417X}}
\email{asal68826@hbku.edu.qa}
\affiliation{Qatar Center for Quantum Computing, College of Science and
Engineering, Hamad Bin Khalifa University, Doha, Qatar}

\author{Saif Al-Kuwari\,\orcidlink{0000-0002-4402-7710}}
\email{smalkuwari@hbku.edu.qa}
\affiliation{Qatar Center for Quantum Computing, College of Science and
Engineering, Hamad Bin Khalifa University, Doha, Qatar}

\date{\today}

\begin{abstract} 
Large momentum transfer (LMT) enhances the gravitational phase of a light-pulse atom interferometer by a factor $n$, while mirrorless operation with momentum-resolved detection can quadruple the phase-carrying quantum Fisher information. These gains compete in practice because the momentum-space fringe period scales as $1/n$, making the fringe signal increasingly vulnerable to finite detector resolution. We analyze this trade-off in a solvable model with instantaneous lossless $n$-photon pulses, a Gaussian source, and Gaussian detection blur, allowing continuous interpolation between Kasevich--Chu and mirrorless geometries. Closed-form expressions for the blurred output distributions and classical Fisher information are obtained, with a fringe-phase averaging approximation whose error is exponentially suppressed and agrees with numerical simulations at the $10^{-8}$ level. We find that mirrorless operation surpasses a conventional interferometer with the same momentum transfer only when $\sigma_p < 0.91\,m/(n k_0 T)$. Fringe-based readout exhibits an optimal momentum transfer $n^* \simeq 0.93\,m/(\sigma_p k_0 T)$ and a resolution-limited sensitivity floor $\Delta g \simeq 4.4\,\sigma_p/(mT\sqrt{N})$, independent of photon momentum. When this criterion is not satisfied, partial mirror asymmetry can recover part of the enhancement, whereas population-based readout remains insensitive to detector blur and ultimately favors the conventional sequence at sufficiently large $n$. Estimates for $^{87}$Rb sensors show that the mirrorless advantage is primarily restricted to short-baseline instruments.
\end{abstract}

\maketitle

\section{Introduction}
\label{sec:intro}

Light-pulse atom interferometers are among the most precise absolute
gravimeters available \cite{Kasevich1991,Peters1999,Peters2001,Bongs2019}.
In the standard Kasevich--Chu (KC) Mach--Zehnder sequence---a
$\pi/2$--$\pi$--$\pi/2$ train of two-photon transitions separated by a time
$T_\pi$---the gravitational acceleration $g$ imprints the phase $\phi = \kO g
T_\pi^2$ on the atomic population difference, where $\hbar\kO$ is the
photon-recoil momentum \cite{Kasevich1991,Storey1994}. Since the achievable
interrogation time is bounded by apparatus size and coherence, considerable
effort has gone into increasing the momentum splitting between the
interferometer arms using large-momentum-transfer (LMT) beam splitters based
on high-order Bragg diffraction, sequential pulses, or Bloch oscillations
\cite{Muller2008,Clade2009,Chiow2011,McDonald2013,Kovachy2015,Ahlers2016},
which multiply the phase, and hence the per-shot sensitivity, by the number
$n$ of transferred photon momenta. Current implementations reach momentum
separations of $141\,\hbar\kO$ on optical clock transitions
\cite{Rudolph2020}, $408\,\hbar\kO$ in twin-lattice geometries
\cite{Gebbe2021}, and $600\,\hbar\kO$ with Floquet-engineered beam splitters
\cite{Rodzinka2024}. The question of how much of the amplified signal
survives the readout therefore becomes increasingly relevant.

A complementary line of work asks, for a \emph{given} motional state and
interrogation time, how much information about $g$ the full quantum state
carries and which measurement extracts it. Using the quantum and classical
Fisher information (QFI and CFI) \cite{Helstrom1976,Braunstein1994},
Kritsotakis \emph{et al.}\ \cite{Kritsotakis2018} showed that the
conventional KC gravimeter is suboptimal in two distinct respects. First, the
population-difference measurement discards correlations between the atomic
momentum and the accumulated phase; measuring the momentum distribution of
each output port recovers them. Second, the mirror pulse itself carries an
information cost: removing it---leaving a Ramsey-type sequence of two $\pi/2$
pulses separated by $T = 2T_\pi$---quadruples the phase-carrying QFI from
$\kO^2 T_\pi^4$ to $4\kO^2 T_\pi^4$, at the price of storing the information
in momentum-space interference fringes whose readout requires high detector
resolution.

These two directions---LMT and measurement optimality---have developed
independently, and their combination is nontrivial. The mirrorless
enhancement multiplies the coefficient of the $\kO^2$ term of the QFI, so at
the level of the QFI it compounds with the $n^2$ gain of LMT. The fringes
that carry this information, however, arise from the interference of wave
packets separated by $\Delta z = n\hbar\kO T/m$, so their momentum-space
period $2\pi\hbar/\Delta z$ decreases linearly in $n$ while the absolute
resolution $\sigma_p$ of a given detector does not. Increasing the momentum
transfer thus amplifies the signal and simultaneously compresses the feature
that must be resolved. The purpose of this paper is to quantify this
competition within a solvable model and to extract the design rules it
implies.

Both ingredients of the analysis have experimental counterparts, though they
have not previously been combined in a Fisher-information treatment.
Momentum-resolved readout of a Bragg Mach--Zehnder gravimeter has been
demonstrated by Raman spectroscopy of the output ports \cite{Cheng2018};
deliberately \emph{asymmetric} pulse timing has been used to generate
overlapped spatial fringes that are imaged directly \cite{Wigley2019}, in
the same spirit as phase-shear and point-source readout, where an imposed
phase gradient converts the interferometer phase into a spatially resolvable
fringe pattern \cite{Sugarbaker2013,Dickerson2013}, and as contrast
interferometry, which reads recoil phases from momentum-space fringes
\cite{Gupta2002}. On the theory side, the mirrorless configuration has been
adopted as a platform for multiparameter decorrelation \cite{Alam2024}. None
of these works optimizes the momentum transfer against detection resolution,
provides a closed-form Fisher information including the detection-blur
contrast penalty, or treats the mirror timing as a continuous
information-theoretic design parameter; the present analysis is, to our
knowledge, the first to treat these three elements jointly, though it does
so within an idealized model whose limitations are discussed in
Sec.~\ref{sec:discussion}. The mirrorless momentum-resolved gravimeter
itself has, to our knowledge, not been realized experimentally; the
detector requirements derived below indicate what such a realization would
demand.

Our results are as follows. (i)~We solve the pulsed dynamics in closed
form at the propagator level and generalize the QFI of Ref.~\cite{Kritsotakis2018} to
$n$-photon momentum transfer and arbitrary mirror-pulse timing, obtaining a
one-parameter family of interferometers that interpolates continuously
between the KC and mirrorless configurations (Sec.~\ref{sec:model};
Appendices \ref{app:dynamics} and \ref{app:qfi}). (ii)~We derive the blurred
output distributions and the CFI of momentum-resolved detection with finite
Gaussian resolution $\sigma_p$ in closed form, the only approximation
being a fringe-phase average whose error is exponentially small in the
squared fringe number; the result agrees with direct numerics at the level
of their precision and reduces, for $\sigma_p \ll \delta p$, to a leading
form containing the fringe-contrast factor and the phase-averaged reduction
factor $1 - \sqrt{1-C^2}$ (Sec.~\ref{sec:cfi};
Appendices \ref{app:ports} and \ref{app:cfi}). (iii)~Maximization over $n$
gives an optimal momentum transfer $n^* \simeq 0.93\,m/(\sigma_p\kO T)$, an
information ceiling $F \le h^* m^2 T^2/(4\sigma_p^2)$ with $h^* \simeq
0.207$ that applies to fringe-based readout but not to population readout,
and a criterion, $\sigma_p \lesssim 0.91\,m/(n\kO T)$, for the mirrorless
scheme to outperform a conventional KC interferometer of equal momentum
transfer (Sec.~\ref{sec:optimal}; Appendix~\ref{app:opt}). (iv)~When the
detector resolution violates this criterion, a partial asymmetry of the
mirror timing, optimized through the same expression, recovers part of the
intermediate gain (Sec.~\ref{sec:asym}; Appendix~\ref{app:asym}). We
conclude with quantitative estimates indicating that mirrorless LMT is best
suited to compact, short-baseline sensors, while long-baseline instruments
are better served by conventional LMT or by slight timing asymmetry
(Sec.~\ref{sec:discussion}). Numerical methods and validation protocols are
collected in Appendix~\ref{app:numerics}.

\section{Model and quantum Fisher information}
\label{sec:model}

We consider a single atom (or a dilute, effectively noninteracting ensemble
of $N$ atoms) of mass $m$ with two internal states $|a\rangle$, $|b\rangle$,
subject to a uniform gravitational acceleration $g$ along $z$, so that the
motional Hamiltonian during free evolution is
$\op{H} = \op{p}_z^2/2m + m g \op{z}$. Beam-splitting pulses couple the
internal states while transferring $n$ photon momenta,
\begin{equation}
\op{U}_\theta = \cos\tfrac{\theta}{2}
- i \left( e^{i n \kO \op{z}} |b\rangle\langle a|
+ e^{-i n \kO \op{z}} |a\rangle\langle b| \right) \sin\tfrac{\theta}{2},
\label{eq:pulse}
\end{equation}
with $\theta = \pi/2$ for beam splitters and $\theta = \pi$ for the mirror.
Equation~\eqref{eq:pulse} idealizes an $n$-photon LMT beam splitter (e.g.,
high-order Bragg diffraction or a composite pulse) as instantaneous and
lossless; the limitations of this idealization are discussed in
Sec.~\ref{sec:discussion}. Laser phases have been set to zero: they carry no
$g$ dependence and, except in the fully symmetric limit discussed in
Appendix~\ref{app:numerics}, the momentum-resolved Fisher information below
is independent of them. The interferometer family under study is
\begin{equation}
\tfrac{\pi}{2} \;\longrightarrow\; T_1 \;\longrightarrow\; \pi
\;\longrightarrow\; T_2 \;\longrightarrow\; \tfrac{\pi}{2},
\qquad T_1 + T_2 = T,
\label{eq:sequence}
\end{equation}
parametrized by the mirror asymmetry
\begin{equation}
s \equiv \tfrac{1}{2}(T_1 - T_2) \in [0, T/2].
\label{eq:asymdef}
\end{equation}
Here $s=0$ is the symmetric KC configuration, and $s = T/2$ (i.e., $T_2 =
0$, where the $\pi$ pulse merges with the final beam splitter into a single
composite pulse, equivalent up to a relabeling of the output ports to the
two-pulse sequence) is the mirrorless configuration of
Ref.~\cite{Kritsotakis2018}. Throughout, configurations are compared at
fixed total interrogation time $T = 2T_\pi$. The initial state is $|a\rangle
\otimes |\psi_0\rangle$ with $|\psi_0\rangle$ a Gaussian wave packet of
position variance $\var(\op z) = \sigma^2/2$ and momentum variance
$\var(\op p_z) = \hbar^2/(2\sigma^2) \equiv \delta p^2$; its
momentum-representation wave function is
$\psi_0(p) = (\sigma^2/\pi\hbar^2)^{1/4} e^{-\sigma^2 p^2/2\hbar^2}$.
The dynamics generated by $\op H$ and the pulses \eqref{eq:pulse} admits a
closed-form solution: the momentum-space propagator and the composition of
the full sequence, including the four branch amplitudes of the output state
and their accumulated phases, are derived in Appendix~\ref{app:dynamics}.

For a pure state $|\psi_g\rangle$ depending on the parameter $g$, the QFI is
\begin{equation}
F_Q = 4\left[ \langle \partial_g \psi_g | \partial_g \psi_g \rangle
- \left| \langle \psi_g | \partial_g \psi_g \rangle \right|^2 \right],
\label{eq:qfidef}
\end{equation}
a formula derived from the fidelity expansion in
Appendix~\ref{app:qfi}\,\ref{app:qfi-def}, and the Cram\'er--Rao bound for
$N$ independent atoms reads $\Delta g \ge 1/\sqrt{N F}$, with $F$ the QFI
for the optimal measurement or the CFI for a specified one
\cite{Helstrom1976,Braunstein1994}. For the sequence \eqref{eq:sequence} the
QFI is evaluated in Appendix~\ref{app:qfi} by two independent routes---a
generator computation using Heisenberg-picture conjugation identities, and
a direct moment computation from the branch amplitudes of
Appendix~\ref{app:dynamics}, which agree---with the result
\begin{equation}
F_Q = 4 \var\!\big[\op{G}_0(T)\big]
+ \frac{n^2 \kO^2}{4}\left( T^2 - 2 T_2^2 \right)^2 ,
\label{eq:QFI}
\end{equation}
where $\op{G}_0(T) = \tfrac{1}{\hbar}\big( mT\op{z} + \tfrac{T^2}{2}
\op{p}_z\big)$ generates the $g$ dependence of the motional state alone, so
that $4\var[\op G_0] = 4(mT/\hbar)^2\var(\op z) + (T^2/\hbar)^2\var(\op
p_z)$ for our state, which has $\mathrm{Cov}(\op z, \op p_z) = 0$. Equation~\eqref{eq:QFI} generalizes Eq.~(24) of
Ref.~\cite{Kritsotakis2018} to $n$-photon kicks and arbitrary timing. The
second (phase) term is the part addressable by interferometry. It reproduces
the known limits: for $s=0$ it gives the semiclassical KC value $n^2\kO^2
T_\pi^4$ with the standard fringe phase $\phi_g = n\kO g T_\pi^2$; for $s =
T/2$ it gives $n^2\kO^2 T^4/4 = 4n^2\kO^2 T_\pi^4$, the fourfold enhancement
of Ref.~\cite{Kritsotakis2018} multiplied by $n^2$; and in general the
fringe phase is
\begin{equation}
\phi_g = \frac{n\kO g}{2}\left( T^2 - 2T_2^2 \right)
= n\kO g \left( T_\pi^2 + 2 T_\pi s - s^2 \right),
\label{eq:phase}
\end{equation}
a consequence of the branch phases derived in
Appendix~\ref{app:dynamics}\,\ref{app:branchphases}.
Equation~\eqref{eq:QFI}, including the motional term and the vanishing of
the motional--internal cross covariance, was additionally validated
numerically through the fidelity of simulated output states
(Appendix~\ref{app:numerics}).

Two structural observations organize the analysis that follows. First, the
motional term $4\var[\op G_0]$ is independent of $\kO$: LMT does not amplify
it, so its relative contribution is diluted as $1/n^2$ and, for the
parameters considered here, it is a small additive correction. Second, the
mirrorless enhancement is a prefactor of the $\kO^2$ term itself, so at the
level of the QFI the two strategies multiply: the mirrorless $n$-photon
interferometer carries $4n^2$ times the phase information of a
single-photon KC device. Whether this information is extractable is a
question about the measurement, to which we now turn.

\section{Momentum-resolved detection at finite resolution}
\label{sec:cfi}

\begin{figure}[t]
\includegraphics[width=\columnwidth]{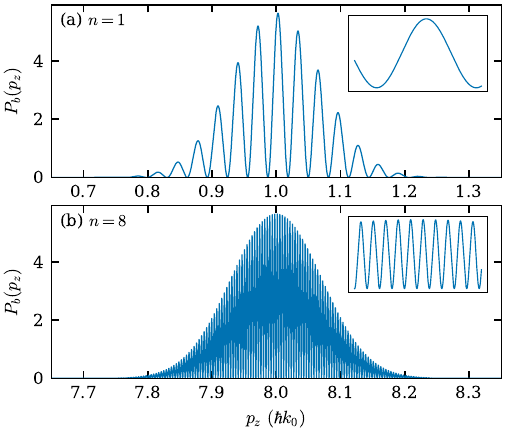}
\caption{Momentum distribution of output port $b$ of the mirrorless
interferometer ($s = T/2$) at $g = 0$, for momentum transfer (a) $n = 1$ and
(b) $n = 8$, with $\sigma = 10\,\kO^{-1}$ and $T = 200\,t_0$, where $t_0 =
m/(\hbar\kO^2)$. Insets: magnified views of the fringes over a window of
width $0.04\,\hbar\kO$. The fringe period $2\pi m/(n\kO T)$ decreases as
$1/n$ while the Gaussian envelope, of width $\delta p = \hbar/(\sqrt{2}\,
\sigma) \approx 0.07\,\hbar\kO$, is unchanged: increasing the momentum
transfer amplifies the phase but compresses the feature that must be
resolved.}
\label{fig:profiles}
\end{figure}

Following Ref.~\cite{Kritsotakis2018}, the measurement is the joint readout
of the internal state and the momentum of each output port, with the
detector modeled by a normalized Gaussian kernel
$G_{\sigma_p}(p) = e^{-p^2/2\sigma_p^2}/\sqrt{2\pi\sigma_p^2}$
of standard deviation $\sigma_p$ in absolute momentum units: the observed
distributions are the convolutions $\tilde P_s = P_s * G_{\sigma_p}$, $s \in
\{a, b\}$, and the CFI is
\begin{equation}
F_C = \sum_{s=a,b} \int dp\,
\frac{\big[\partial_g \tilde P_s(p)\big]^2}{\tilde P_s(p)} .
\label{eq:cfidef}
\end{equation}

The structure of the output distributions follows in
Appendix~\ref{app:ports} from the branch amplitudes of
Appendix~\ref{app:dynamics}. In each port, the final state is a
superposition of two wave packets whose spatial separation at the final
beam splitter is
\begin{equation}
\Delta z = \frac{n \hbar \kO}{m} (T_1 - T_2) = \frac{2 n \hbar \kO s}{m},
\label{eq:dz}
\end{equation}
equal to $n\hbar\kO T/m$ in the mirrorless limit. The two interfering
amplitudes have identical momentum-space envelopes and a
relative phase that is linear in $p$ without approximation
[Eqs.~\eqref{eq:relphasea} and \eqref{eq:relphaseb}], so the port
distributions are
\begin{equation}
P_{a,b}(p) = \tfrac12 Q(p + mgT) \left[ 1 \pm \cos\!\big( \kappa p +
\phi_g \pm \phi_r \big) \right],
\label{eq:fringes}
\end{equation}
where $\kappa \equiv \Delta z/\hbar$ is the fringe wave number, $Q$ is the
common Gaussian envelope of variance $\delta p^2$ (centered at $0$ for port
$a$ and at $n\hbar\kO$ for port $b$; the shift is left implicit), $\phi_g$
is the fringe phase \eqref{eq:phase}, and $\phi_r = n^2\hbar\kO^2
(T_1-T_2)/2m$ is a $g$-independent recoil phase. The intrinsic fringe
contrast is unity (Fig.~\ref{fig:profiles}), and the entire $g$
dependence resides in the rigid envelope shift $mgT$ and in $\phi_g$.

The convolution of Eq.~\eqref{eq:fringes} with the Gaussian kernel can be
carried out in closed form (Appendix~\ref{app:cfi}\,\ref{app:conv}).
Defining
\begin{equation}
\Sig^2 \equiv \delta p^2 + \sigma_p^2, \quad
\kappa_e \equiv \frac{\kappa\,\delta p^2}{\Sig^2}, \quad
C_e \equiv e^{-\kappa^2 \sigma_p^2 \delta p^2 / 2\Sig^2},
\label{eq:blurdefs}
\end{equation}
the blurred port distributions are
\begin{equation}
\tilde P_{a,b}(p) = \tfrac12 Q_\Sig(p + mgT)\left[ 1 \pm C_e \cos\!\Big(
\kappa_e p + \tilde\phi_g \pm \phi_r \Big) \right],
\label{eq:blurred}
\end{equation}
with $Q_\Sig$ the Gaussian of variance $\Sig^2$ and
\begin{equation}
\tilde\phi_g = \phi_g - \kappa\, m g T\, \frac{\sigma_p^2}{\Sig^2},
\quad
\partial_g\tilde\phi_g = \partial_g\phi_g
- \frac{\kappa m T \sigma_p^2}{\Sig^2}.
\label{eq:phitilde}
\end{equation}
Blurring thus does three things to the fringe pattern: it broadens the envelope ($\delta p^2 \to \Sig^2$), it reduces
the contrast ($1 \to C_e$) and stretches the fringe period ($\kappa \to
\kappa_e$), and---less obviously---it shifts the effective fringe phase by a
$g$-dependent amount, because the blurred fringe pattern partially tracks
the moving envelope.

Inserting Eq.~\eqref{eq:blurred} into Eq.~\eqref{eq:cfidef} and summing the
two ports, the fringe--envelope cross terms cancel between the ports without
approximation, the envelope terms combine to the Fisher information of
locating a Gaussian of variance $\Sig^2$, and the fringe terms reduce to a
single average over the fringe phase, evaluated in
Appendix~\ref{app:cfi}\,\ref{app:avg}:
\begin{equation}
\left\langle \frac{C_e^2 \sin^2\alpha}{1 - C_e^2 \cos^2\alpha}
\right\rangle_{\!\alpha} = 1 - \sqrt{1 - C_e^2} \;\equiv\; R(C_e^2).
\label{eq:R}
\end{equation}
Combining the three contributions gives
\begin{equation}
F_C = \big( \partial_g \tilde\phi_g \big)^2\, R(C_e^2)
\;+\; \frac{m^2 T^2}{\Sig^2}
\label{eq:master}
\end{equation}
with $\partial_g\tilde\phi_g$, $C_e$, and $\Sig$ given by
Eqs.~\eqref{eq:blurdefs} and \eqref{eq:phitilde}. Within the Gaussian
model, Eq.~\eqref{eq:master} involves one approximation: the replacement of
the envelope-weighted fringe-phase average by the uniform average
\eqref{eq:R}, whose relative error is bounded by
$K(C_e)\,e^{-2\kappa_e^2\Sig^2}$ at leading harmonic, with $K$ of order
unity for the contrasts considered here
(Appendix~\ref{app:cfi}\,\ref{app:error}); the exponential factor is below
$e^{-69}$ for all configurations shown in the figures and below $e^{-398}$
for the entries of Table~\ref{tab:validation}. Consistently, direct
numerical simulation of the full pulse sequence
(Appendix~\ref{app:numerics}) agrees with Eq.~\eqref{eq:master} to a
relative accuracy of $10^{-8}$, limited by the finite-difference step of
the numerics, across mirrorless, symmetric, and intermediate-asymmetry
configurations (Table~\ref{tab:validation}). We emphasize that
Eq.~\eqref{eq:master} is a statement about the model of
Sec.~\ref{sec:model}---instantaneous lossless pulses, a pure Gaussian
state, and a Gaussian detection kernel---not about any particular
apparatus; Sec.~\ref{sec:discussion} discusses which of its features
should survive the idealizations.

For $\sigma_p \ll \delta p$, Eq.~\eqref{eq:master} reduces to the leading
form
\begin{equation}
F_C \simeq \frac{n^2 \kO^2}{4} \big( T^2 - 2T_2^2 \big)^2\, R(C^2)
+ \frac{m^2 T^2}{\delta p^2 + \sigma_p^2},
\label{eq:leading}
\end{equation}
with $C = e^{-\sigma_p^2\Delta z^2/2\hbar^2}$ the leading-order contrast and
relative corrections of order $\sigma_p^2/\delta p^2$
(Appendix~\ref{app:cfi}\,\ref{app:leading}); the design analysis of
Secs.~\ref{sec:optimal} and \ref{sec:asym} is carried out with
Eq.~\eqref{eq:leading} and inherits this accuracy. The first term of
Eq.~\eqref{eq:leading} is the fringe (phase) information: the full QFI
phase term reduced by $R(C^2)$, which interpolates from $R = 1$ at unit
contrast to $R \simeq C^2/2$ when the fringes are nearly washed out; the
contrast exponent grows as $n^2$ at fixed $\sigma_p$, which is the
quantitative content of the LMT--readout competition. The second term is
the envelope information: the gravitational momentum kick $mgT$ displaces
the port distributions rigidly, and its Fisher information is that of
locating a Gaussian of total variance $\delta p^2 + \sigma_p^2$. [The
remaining motional information, the $(T^2/\hbar)^2\var(\op p_z)$ piece of
$4\var[\op G_0]$, is generated by the $\op p_z$ part of $\op G_0$, which
displaces the wave packet in \emph{position} by $gT^2/2$, and is therefore
invisible to a momentum-basis measurement; for a minimum-uncertainty
Gaussian the captured envelope information equals the position-variance
term, $m^2T^2/\delta p^2 = 4(mT/\hbar)^2\var(\op z)$. Both identifications
were verified numerically at parameters that lift the accidental degeneracy
of our default parameter set, for which the two motional pieces coincide
(Appendix~\ref{app:numerics}). For the present parameters the invisible
piece is a $2\%$ effect at $n=1$ and negligible beyond.]

\begin{table}[b]
\caption{Validation of the closed-form CFI against direct numerical
simulation of the full pulse sequence (Appendix~\ref{app:numerics}), in
units of $\kO^2 T_\pi^4$, for $\sigma = 10\,\kO^{-1}$, $T = 200\,t_0$. The
upper block is the mirrorless configuration $s = T_\pi$; the lower block has
intermediate mirror asymmetry. The closed form, Eq.~\eqref{eq:master},
matches the numerics to all digits shown (relative deviations
$\lesssim 10^{-8}$, set by the finite-difference step); the leading form,
Eq.~\eqref{eq:leading}, deviates by $O(\sigma_p^2/\delta p^2)$, at most
$1\%$ here.}
\label{tab:validation}
\begin{ruledtabular}
\begin{tabular}{cccccc}
$n$ & $s/T_\pi$ & $\sigma_p\,(\hbar\kO)$ & numerics &
Eq.~\eqref{eq:master} & Eq.~\eqref{eq:leading} \\
\hline
1  & 1    & $4\times10^{-3}$ & 1.31680   & 1.31680   & 1.3296 \\
3  & 1    & $3\times10^{-3}$ & 0.79091   & 0.79091   & 0.7918 \\
5  & 1    & $1\times10^{-3}$ & 20.56221  & 20.56221  & 20.5740 \\
10 & 1    & $5\times10^{-4}$ & 82.04419  & 82.04419  & 82.0560 \\
15 & 1    & $3\times10^{-4}$ & 229.49677 & 229.49677 & 229.5094 \\
5  & 1    & $3\times10^{-2}$ & 0.06780   & 0.06780   & 0.0678 \\
\hline
5  & 0.5  & $1\times10^{-3}$ & 40.61838  & 40.61838  & 40.6337 \\
5  & 0.25 & $2\times10^{-3}$ & 27.42150  & 27.42150  & 27.4434 \\
10 & 0.1  & $1\times10^{-3}$ & 113.63634 & 113.63634 & 113.6489 \\
8  & 0.4  & $5\times10^{-4}$ & 118.50161 & 118.50161 & 118.5119 \\
\end{tabular}
\end{ruledtabular}
\end{table}

In the limit $\sigma_p \to 0$ both forms reduce to $F_C =
\tfrac14 n^2\kO^2(T^2 - 2T_2^2)^2 + m^2T^2/\delta p^2$, confirming that
momentum-resolved detection extracts the entire phase QFI together with the
wave-packet-locating (position-variance) part of the motional QFI,
consistent with the $n=1$ findings of Ref.~\cite{Kritsotakis2018}.

\section{Optimal momentum transfer and the resolution ceiling}
\label{sec:optimal}

\begin{figure}[t]
\includegraphics[width=\columnwidth]{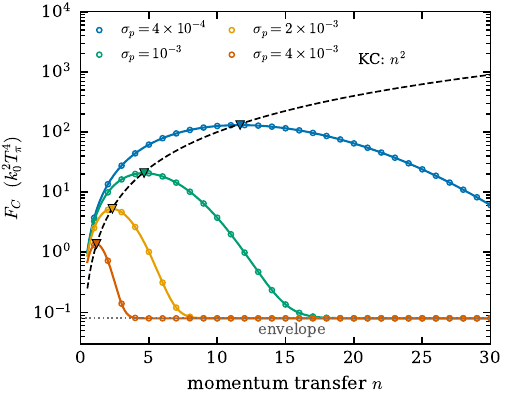}
\caption{CFI of the mirrorless interferometer versus momentum transfer $n$
for four detector resolutions $\sigma_p$ (in units of $\hbar\kO$; legend).
Solid lines: closed form, Eq.~\eqref{eq:master}; circles: direct
numerics; triangles: the optimum $n^*$ of Eq.~\eqref{eq:nstar}. Dashed
line: conventional KC interferometer with the same momentum transfer, $F =
n^2 \kO^2 T_\pi^4$, which is insensitive to $\sigma_p$
(Sec.~\ref{sec:optimal}). Dotted line: envelope information floor.
Parameters as in Fig.~\ref{fig:profiles}.}
\label{fig:Fvsn}
\end{figure}

Consider first the mirrorless configuration, $s = T_\pi$, in the design
regime $\sigma_p \ll \delta p$ where the leading form \eqref{eq:leading}
applies. Its fringe term depends on $n$ only through
\begin{equation}
a \equiv \frac{\sigma_p^2 n^2 \kO^2 T^2}{m^2}
= \left( \frac{\sigma_p \Delta z}{\hbar} \right)^{\!2},
\qquad
F_{\rm fringe} = \frac{m^2 T^2}{4\sigma_p^2}\, h(a),
\label{eq:hdef}
\end{equation}
with $h(a) = a\,[1 - \sqrt{1 - e^{-a}}\,]$, so the CFI has an interior
maximum in $n$ (Fig.~\ref{fig:Fvsn}). As shown in
Appendix~\ref{app:opt}\,\ref{app:fixedpoint}, setting $h'(a) = 0$ and
writing $w \equiv \sqrt{1-e^{-a}}$ reduces the stationarity condition to the
fixed-point equation
\begin{equation}
a = \frac{2w}{1+w}, \qquad w = \sqrt{1 - e^{-a}},
\label{eq:fixedpoint}
\end{equation}
whose unique positive solution is $a^* = 0.86404$, $w^* = 0.76062$, hence
$R(a^*) = 1 - w^* = 0.23938$ and $h^* \equiv h(a^*) = 0.20683$, with
$h''(a^*) = -0.2275 < 0$ confirming the maximum. The optimal momentum
transfer is therefore
\begin{equation}
n^* = \sqrt{a^*}\, \frac{m}{\sigma_p \kO T}
= 0.9295\, \frac{m}{\sigma_p \kO T},
\label{eq:nstar}
\end{equation}
equivalently $\sigma_p \Delta z^*/\hbar = 0.930$: the optimal momentum
transfer places the wave-packet separation at the detector's resolving
length. Beyond $n^*$, additional photon recoils reduce the extractable
information exponentially. At the optimum,
\begin{equation}
F_C(n^*) = \frac{h^*}{4}\, \frac{m^2 T^2}{\sigma_p^2}
= 0.0517\, \frac{m^2 T^2}{\sigma_p^2},
\label{eq:ceiling}
\end{equation}
a resolution-limited ceiling on fringe-based readout that is independent of
$\kO$ and $n$ (population readout is not subject to it; see below). By the
Cram\'er--Rao bound, the corresponding shot-noise-limited sensitivity for
$N$ independent atoms is
\begin{equation}
\Delta g_{\min} = \frac{2\,\sigma_p}{m T \sqrt{N h^*}}
= \frac{4.398\, \sigma_p}{m T \sqrt{N}} .
\label{eq:floor}
\end{equation}
Equation~\eqref{eq:floor} summarizes the readout constraint: once the
momentum transfer is optimized, fringe-based mirrorless gravimetry measures
$g$ with an effective momentum uncertainty set by the detector rather than
by the photon momentum. It has the same form as the envelope-only
sensitivity $\Delta g = \Sig/(mT\sqrt{N})$ with $\Sig \to
2\sigma_p/\sqrt{h^*}$: the fringe structure substitutes the detector
resolution for the quantum momentum width. Comparing the two terms of
Eq.~\eqref{eq:leading} at $n = n^*$
(Appendix~\ref{app:opt}\,\ref{app:crossderive}) shows that the fringe
strategy is advantageous only if
\begin{equation}
\sigma_p < \sqrt{h^*/4}\;\delta p = 0.2274\,\delta p ;
\label{eq:crossover}
\end{equation}
a detector that cannot resolve substantially below the source's own momentum
width gains nothing from the fringes, and the momentum transfer $n$ is then
irrelevant to the attainable information.

\begin{figure}[t]
\includegraphics[width=\columnwidth]{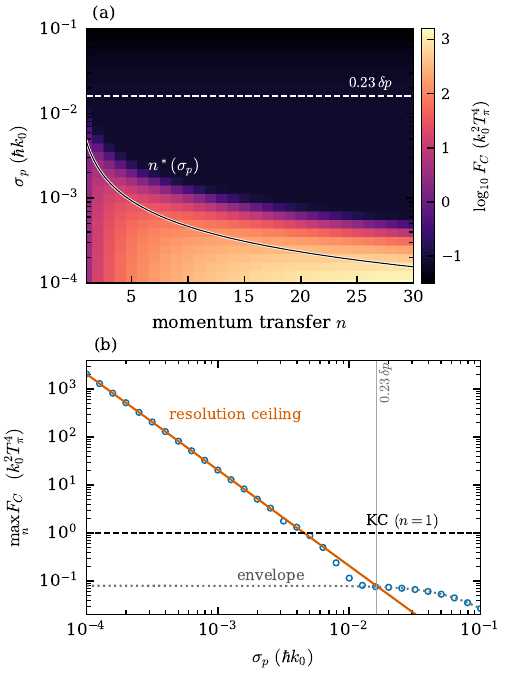}
\caption{(a) CFI of the mirrorless interferometer over momentum transfer
$n$ and detector resolution $\sigma_p$ (direct numerics; color scale:
$\log_{10} F_C$ in units of $\kO^2 T_\pi^4$). The white curve is the
optimal momentum transfer $n^*(\sigma_p)$, Eq.~\eqref{eq:nstar}; the dashed
line marks the fringe--envelope crossover, Eq.~\eqref{eq:crossover}. (b)
CFI at the optimal $n$ (circles: numerical maximum over $n$, evaluated at
$n = \mathrm{round}(n^*)$ where $n^* > 30$) compared with the resolution
ceiling, Eq.~\eqref{eq:ceiling}, the envelope information (dotted), and the
$n{=}1$ KC value (dashed). The numerics follow the ceiling until the
crossover, Eq.~\eqref{eq:crossover} (vertical line), and ride the envelope
floor beyond it.}
\label{fig:optimal}
\end{figure}

\emph{Comparison with conventional LMT.} The appropriate benchmark for
mirrorless LMT is the KC interferometer with the same momentum transfer.
For the symmetric sequence the interfering wave packets overlap at the
final beam splitter [$\Delta z = 0$, Eq.~\eqref{eq:dz}], so the fringe
phase is global rather than momentum dependent: the signal resides in the
port populations, and the $\Delta z \to 0$ limit of the analysis
(Appendix~\ref{app:cfi}\,\ref{app:leading}) gives $F_C = n^2\kO^2 T_\pi^4 +
m^2T^2/\Sig^2$ at the mid-fringe operating point, with no contrast
penalty at any $\sigma_p$. Numerical evaluation confirms this: $F_C$ for the KC
sequence agrees with this expression to within $0.1\%$ as $\sigma_p$ ranges
over three orders of magnitude up to $0.3\,\hbar\kO$, for $n = 1$, $5$, and
$10$. The mirrorless configuration therefore outperforms a KC device of
equal $n$ if and only if $4R(C^2) > 1$, i.e., $C^2 > 7/16$, which
translates (Appendix~\ref{app:opt}\,\ref{app:critderive}) into the
resolution criterion
\begin{equation}
\frac{\sigma_p \Delta z}{\hbar} < \sqrt{\ln(16/7)} = 0.9092
\;\;\Longleftrightarrow\;\;
\sigma_p < 0.91\, \frac{m}{n \kO T}.
\label{eq:criterion}
\end{equation}
Equations~\eqref{eq:nstar} and \eqref{eq:criterion} carry the same content
with different emphasis: mirrorless operation and LMT are compatible up to
$n \sim m/(\sigma_p \kO T)$, where the fourfold enhancement is essentially
intact, and incompatible beyond, where the conventional mirror
sequence---whose signal survives arbitrary detector blurring---is superior.
The two numerical constants nearly coincide, $\sqrt{a^*} = 0.930$ against
the break-even value $0.909$, with one consequence: since $4R(a^*) =
0.958 < 1$, the mirrorless interferometer at its own optimal momentum
transfer is already marginally inferior to a KC device of the same $n$. The
mirrorless advantage is therefore confined to $n$ below the crossover of
Eq.~\eqref{eq:criterion}, where the enhancement factor $4R$ approaches
$4$. The practical case for removing the mirror arises when technical
constraints---pulse fidelity, laser power, diffraction losses---cap the
achievable momentum transfer below $n^*$ and a detector satisfying
Eq.~\eqref{eq:criterion} at that capped $n$ is available. Conversely, when
arbitrarily large $n$ is available, the conventional sequence prevails: its
information grows as $n^2$ without a resolution ceiling.

\section{Optimal mirror asymmetry}
\label{sec:asym}

\begin{figure}[t]
\includegraphics[width=\columnwidth]{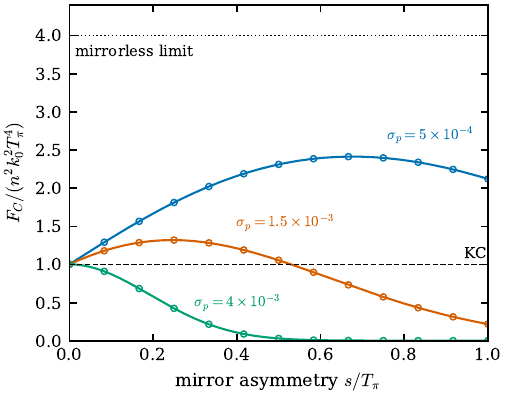}
\caption{CFI, normalized to the KC value $n^2 \kO^2 T_\pi^4$, across the
interpolating family of interferometers, versus mirror asymmetry $s$, for
$n = 5$ and three detector resolutions (labels adjacent to each curve, in
units of $\hbar\kO$). Solid lines: closed form,
Eq.~\eqref{eq:master}; circles: direct numerics. For good resolution the
optimum is the fully mirrorless sequence ($s = T_\pi$); for poorer
resolution an interior optimum $s^*$ emerges, and a partial asymmetry still
exceeds the symmetric KC value (dashed line).}
\label{fig:asym}
\end{figure}

The configurations compared above are the endpoints of the continuous
family of Eq.~\eqref{eq:sequence}, and Eq.~\eqref{eq:master} covers the
interior. Writing $u = s/T_\pi \in [0,1]$, the fringe term of the leading
form is proportional to (Appendix~\ref{app:asym})
\begin{equation}
f(u) = \big( 2 - (1-u)^2 \big)^2\, R\!\left( e^{-\beta u^2} \right),
\qquad
\beta = \frac{4 \sigma_p^2 n^2 \kO^2 T_\pi^2}{m^2},
\label{eq:family}
\end{equation}
normalized so that $f(0) = 1$ (KC) and $f(1) \to 4$ as $\sigma_p \to 0$.
Figure~\ref{fig:asym} shows the family for $n = 5$: the closed form tracks
the direct numerics through the interior optimum. Because $R(e^{-x}) = 1 -
\sqrt{1 - e^{-x}} = 1 - \sqrt{x} + O(x^{3/2})$ for small $x$, the function
$f$ is nonanalytic at $u = 0$, with the one-sided expansion
(Appendix~\ref{app:asym}\,\ref{app:uexp})
\begin{equation}
f(u) = 1 + \big(4 - \sqrt{\beta}\big) u + \big(2 - 4\sqrt{\beta}\big) u^2
+ O\!\big(\beta^{3/2} u^3\big),
\label{eq:fexp}
\end{equation}
so an infinitesimal asymmetry increases the information whenever
$\sqrt{\beta} < 4$, i.e.,
\begin{equation}
\sigma_p < \frac{2m}{n \kO T_\pi},
\label{eq:asymcriterion}
\end{equation}
a condition roughly four times less restrictive than the full-mirrorless
criterion, Eq.~\eqref{eq:criterion}. Mirror-timing asymmetry is thus a
continuous design parameter: it exchanges a controllable fraction of the
mirrorless enhancement for fringe visibility, with the optimum $s^*$
determined by the stationarity condition given in
Appendix~\ref{app:asym}\,\ref{app:ustat}. Near the threshold, $0 < 4 -
\sqrt{\beta} \ll 4$, the maximum of Eq.~\eqref{eq:fexp} lies at
\begin{equation}
u^* = \frac{4 - \sqrt{\beta}}{8\sqrt{\beta} - 4} + O\big((4 -
\sqrt{\beta})^2\big),
\label{eq:ustar}
\end{equation}
at which the separation remains below the detector's resolving length,
$\sigma_p\Delta z^*/\hbar = \sqrt{\beta}\,u^* < 1$
(Appendix~\ref{app:asym}\,\ref{app:uexp}).

The family has direct experimental precedent. Asymmetric-timing Bragg
interferometry with overlapped, directly imaged spatial fringes has been
demonstrated \cite{Wigley2019}, and phase-shear readout deliberately imposes
a phase gradient across the cloud for the same purpose
\cite{Sugarbaker2013}. The $\Delta z$ scaling underlying the contrast
factor has also been observed directly: in twin-lattice interferometry the
output contrast decays with the wave-packet displacement $\delta T \times
K$ when the second free-evolution time is detuned by $\delta T$ at relative
momentum $K$, with an envelope width shrinking as $1/K$
\cite{Gebbe2021}---the same Gaussian-overlap mechanism, arising there from
the finite spatial coherence length of the source rather than from detector
blur. Those experiments chose their asymmetry (or shear) empirically to
balance fringe visibility against signal; Eqs.~\eqref{eq:master} and
\eqref{eq:family} provide the Fisher-information statement of that
trade-off and its optimum.

\section{Experimental estimates and discussion}
\label{sec:discussion}

To attach laboratory numbers to these scalings, consider $^{87}$Rb ($m =
1.443\times10^{-25}$~kg) with a two-photon effective wave vector $\kO =
1.61\times10^{7}\,$m$^{-1}$ ($\hbar\kO = 1.70\times10^{-27}$~kg\,m/s), so
that $t_0 = m/(\hbar\kO^2) = 5.3\,\mu$s. Momentum widths quoted below are
in units of $\hbar\kO$. Absolute momentum resolution of order
$10^{-2}\,\hbar\kO$ is routine; reaching $10^{-4}\,\hbar\kO$ requires
box-trapped or delta-kick-collimated condensates with long expansion, for
which momentum spreads far below a single recoil have been observed
\cite{Gotlibovych2014}, or spectroscopic velocimetry
\cite{Stenger1999,Debs2011}; momentum-resolved readout of a Bragg
gravimeter has been demonstrated with Raman spectroscopy of the output
ports \cite{Cheng2018}.

\emph{Long-baseline instruments.} For an interrogation time $T = 2T_\pi =
260$~ms \cite{Hardman2016}, $T = 4.9\times10^{4}\,t_0$ and
Eq.~\eqref{eq:criterion} requires $\sigma_p < 1.9\times10^{-5}/n$: even the
$n{=}1$ mirrorless advantage demands resolution an order of magnitude
beyond current capability, and LMT tightens the requirement further.
Long-$T$ instruments should therefore use conventional LMT, for which the
$n^2$ gain is robust to detection---or a slight timing asymmetry:
Eq.~\eqref{eq:asymcriterion} is satisfied at $n = 1$ for $\sigma_p \lesssim
8\times10^{-5}$, and at $\sigma_p = 2\times10^{-5}$ the optimal asymmetry
$s^* \approx 0.4\,T_\pi$ yields a factor $\approx 1.7$ over KC at equal
$T$.

\emph{Compact sensors.} The situation is reversed for short baselines.
With $T = 10$~ms ($1.9\times10^{3}\,t_0$) and $\sigma_p = 10^{-4}$,
Eq.~\eqref{eq:nstar} gives $n^* \approx 5$: the mirrorless scheme
outperforms an equal-$n$ KC device for all $n \le 4$
[Eq.~\eqref{eq:criterion}], with enhancement factor $4R$ decreasing from
$3.2$ at $n=1$ to $1.4$ at $n=4$, and the optimized information ceiling,
Eq.~\eqref{eq:ceiling}, is approximately $23$ times the $n{=}1$ KC value.
Because $n^* \propto 1/T$, mirrorless LMT is best suited to compact
sensors: the shorter the interrogation time, the more momentum transfer the
detector can afford to read out. The resolution floor,
Eq.~\eqref{eq:floor}, is $\Delta g\sqrt{N} \approx
5\times10^{-4}\,$m\,s$^{-2}$ for these numbers, improving linearly with
detector resolution rather than with photon momentum.

The scope of these results is bounded by the idealizations of the model,
which we now discuss at the same level of detail as the results themselves.
First, the readout:
$\sigma_p$ is not a free parameter independent of the scheme, because the
scheme generates its own resolution requirement. Time-of-flight imaging of
two coherent wave packets separated by $\Delta z$ displays clean momentum
fringes only in the far field, $t_{\rm tof} \gg m\sigma_z\Delta z/\hbar$
with $\sigma_z$ the source size, which for the compact-sensor example above
($\Delta z \approx 0.6$~mm) implies expansion times approaching a second
even for a submicron source, and proportionally longer for realistic
micron-scale clouds; the natural readouts for this scheme are instead
spectroscopic---Bragg or Raman velocimetry of the output ports, as
demonstrated in Ref.~\cite{Cheng2018}---whose resolution is set by pulse
duration rather than by expansion. Second, we modeled the LMT beam splitter
as a single instantaneous $n$-photon two-port coupling,
Eq.~\eqref{eq:pulse}. Real high-order Bragg pulses have a velocity
acceptance that decreases with diffraction order
\cite{Szigeti2012,MullerBragg2008}, filtering the very momentum
distribution whose width $\delta p$ enters the envelope term; Bragg
diffraction also leaves the internal state unchanged, so that the output
ports are labeled by momentum alone, folding port identification into the
same resolution budget. Sequential-pulse and Bloch-oscillation LMT
introduce additional timing structure (and losses) whose
Fisher-information cost can be computed with the same tools, since the
generator algebra behind Eq.~\eqref{eq:QFI} accommodates arbitrary pulse
trains (Appendix~\ref{app:qfi}); the optimization over pulse-train timings
is left to future work.

Third, the analysis assumes a pure Gaussian motional state and neglects
interatomic interactions---a relevant caveat because the narrow $\delta p$
favoring the envelope term points toward condensed sources---as well as
finite pulse duration; the Gaussian detection kernel itself is a model, and
real detectors add binning, background, and nonlinearity that we have not
treated. Fourth, the Fisher information bounds the asymptotic sensitivity of
an unbiased estimator over many shots; attaining it requires an estimator
that uses the full momentum-resolved record, and single-shot or
small-sample performance may fall short of the Cram\'er--Rao value.
Moreover, the optimization is carried out within momentum-basis
measurements, leaving open whether a different observable is superior at
finite resolution. Finally, LMT amplifies sensitivity to gravity gradients
and rotations along with the signal; the systematics of the mirrorless
configuration warrant a dedicated treatment. The central mechanism---the
$1/n$ compression of the information-bearing fringes against a fixed
absolute detector resolution---does not depend on the Gaussian
idealizations and should persist in any implementation, but the numerical
constants derived here will shift accordingly.

In summary, the fourfold mirrorless enhancement of
Ref.~\cite{Kritsotakis2018} and the $n^2$ gain of large momentum transfer
multiply at the level of the quantum Fisher information but are jointly
limited by the detector: within the model studied, the extractable
information is given by Eq.~\eqref{eq:master}, is maximized at the momentum transfer of
Eq.~\eqref{eq:nstar}, which places the arm separation at the detector's
resolving length, and cannot exceed the $\kO$-independent ceiling of
Eq.~\eqref{eq:ceiling}---a ceiling that does not apply to population
readout. The criteria of Eqs.~\eqref{eq:criterion} and
\eqref{eq:asymcriterion} specify, in laboratory quantities, when removal of
the mirror pulse is advantageous, when a partial timing asymmetry is
preferable, and when the conventional Mach--Zehnder remains the better
choice.

\section*{Acknowledgments}
This publication was partially supported by the Qatar Research, Development and Innovation (QRDI) Council under the Academic Research Grant ARG01-0603-230468. The findings and views expressed herein are solely the responsibility of the authors.

\appendix

\section{Closed-form solution of the pulsed dynamics}
\label{app:dynamics}

This appendix solves the model of Sec.~\ref{sec:model} in closed form: the
free-fall
propagator in the momentum representation
(\ref{app:prop}), the action of the pulses (\ref{app:pulseaction}), the
closed-form composition of the full sequence (\ref{app:composition}), and
the accumulated branch phases (\ref{app:branchphases}) from which every
subsequent result follows.

\subsection{Momentum-space propagator}
\label{app:prop}

In the momentum representation, $\op z = i\hbar\,\partial_p$, and the
Schr\"odinger equation $i\hbar\,\partial_t\psi = (\op p_z^2/2m + mg\op
z)\psi$ becomes the first-order partial differential equation
\begin{equation}
\partial_t \psi(p,t) - mg\,\partial_p \psi(p,t)
= -\frac{i p^2}{2m\hbar}\, \psi(p,t).
\label{eq:pde}
\end{equation}
Along the characteristic curves $p(\tau) = p + mg(t-\tau)$, $0 \le \tau \le
t$ (so that $p(t) = p$ and $p(0) = p + mgt$), Eq.~\eqref{eq:pde} reduces to
the ordinary differential equation $d\psi/d\tau = -\tfrac{i}{2m\hbar}
p(\tau)^2\,\psi$, whose solution is
\begin{align}
\psi(p,t) &= \psi(p + mgt,\, 0)\,
\exp\!\left[ -\frac{i}{2m\hbar} \int_0^t \big( p + mg s \big)^2\, ds \right]
\nonumber\\
&= \psi(p + mgt,\, 0)\; e^{-i\Theta(p,t)/\hbar},
\label{eq:prop}
\end{align}
where the substitution $s = t - \tau$ was used and the accumulated phase is
\begin{equation}
\Theta(p,t) = \frac{p^2 t}{2m} + \frac{g\, p\, t^2}{2}
+ \frac{m g^2 t^3}{6},
\label{eq:Theta}
\end{equation}
obtained by expanding the square and integrating term by term:
$\int_0^t (p^2 + 2mgps + m^2g^2s^2)\,ds = p^2 t + mgp t^2 + m^2g^2t^3/3$.
Direct substitution into Eq.~\eqref{eq:pde} verifies
Eq.~\eqref{eq:prop}. Physically, free fall shifts the momentum
argument rigidly by $mgt$ and multiplies by a phase that is quadratic in
$p$ (dispersion), linear in $p$ (the displacement-generating term that will
carry the invisible part of the motional information), and a $p$-independent
c-number.

\subsection{Action of the pulses}
\label{app:pulseaction}

The kick operators act in the momentum representation as boosts,
$\big(e^{\pm i n\kO \op z}\psi\big)(p) = \psi(p \mp n\hbar\kO)$, which
follows from $e^{\pm i n\kO \op z}|p\rangle = |p \pm n\hbar\kO\rangle$.
Writing the two-component motional spinor as $(\psi_a, \psi_b)$, the pulses
\eqref{eq:pulse} act as
\begin{align}
\op U_{\pi/2}:\quad
\psi_a(p) &\to \tfrac{1}{\sqrt2}\big[ \psi_a(p) - i\,\psi_b(p + n\hbar\kO)
\big],
\nonumber\\
\psi_b(p) &\to \tfrac{1}{\sqrt2}\big[ \psi_b(p) - i\,\psi_a(p - n\hbar\kO)
\big],
\label{eq:pihalfaction}\\
\op U_{\pi}:\quad
\psi_a(p) &\to -i\,\psi_b(p + n\hbar\kO),
\nonumber\\
\psi_b(p) &\to -i\,\psi_a(p - n\hbar\kO).
\label{eq:piaction}
\end{align}

\subsection{Composition of the full sequence}
\label{app:composition}

For the composition we employ natural units $\hbar = m = \kO = 1$
(momentum in $\hbar\kO$, time in $t_0 = m/\hbar\kO^2$, length in
$\kO^{-1}$); dimensions are restored in all final results. The initial
state is $\psi_a^{(0)}(p) = \psi_0(p)$, $\psi_b^{(0)} = 0$. Applying
Eq.~\eqref{eq:pihalfaction}, then the propagator \eqref{eq:prop} for time
$T_1$, Eq.~\eqref{eq:piaction}, the propagator for $T_2$, and
Eq.~\eqref{eq:pihalfaction} again, each step in closed form:
\begin{align}
\op U_{\pi/2}\!:\;\;
& A_0(p) = \tfrac{1}{\sqrt2}\psi_0(p), \quad
B_0(p) = -\tfrac{i}{\sqrt2}\psi_0(p - n);
\nonumber\\
U_g(T_1)\!:\;\;
& A_1(p) = A_0(p + gT_1)\, e^{-i\Theta(p,T_1)},
\nonumber\\
& B_1(p) = B_0(p + gT_1)\, e^{-i\Theta(p,T_1)};
\nonumber\\
\op U_{\pi}\!:\;\;
& A_2(p) = -i B_1(p + n), \quad B_2(p) = -i A_1(p - n);
\nonumber\\
U_g(T_2)\!:\;\;
& A_3(p) = A_2(p + gT_2)\, e^{-i\Theta(p,T_2)},
\nonumber\\
& B_3(p) = B_2(p + gT_2)\, e^{-i\Theta(p,T_2)};
\nonumber\\
\op U_{\pi/2}\!:\;\;
& A_{\rm out}(p) = \tfrac{1}{\sqrt2}\big[ A_3(p) - i B_3(p + n) \big],
\nonumber\\
& B_{\rm out}(p) = \tfrac{1}{\sqrt2}\big[ B_3(p) - i A_3(p - n) \big].
\label{eq:steps}
\end{align}
Substituting each line into the next and using $T = T_1 + T_2$, the four
branch amplitudes evaluate to
\begin{align}
A_{\rm out}(p) &= -\tfrac12\, \psi_0(p + gT)\,
\Big[ e^{i\Phi_{a,1}(p)} + e^{i\Phi_{a,2}(p)} \Big],
\label{eq:Aout}\\
B_{\rm out}(p) &= -\tfrac{i}{2}\, \psi_0(p + gT - n)\,
\Big[ e^{i\Phi_{b,1}(p)} - e^{i\Phi_{b,2}(p)} \Big],
\label{eq:Bout}
\end{align}
with the accumulated phases
\begin{align}
\Phi_{a,1} &= -\big[ \Theta(p + gT_2 + n,\, T_1) + \Theta(p,\, T_2) \big],
\nonumber\\
\Phi_{a,2} &= -\big[ \Theta(p + gT_2,\, T_1) + \Theta(p + n,\, T_2) \big],
\nonumber\\
\Phi_{b,1} &= -\big[ \Theta(p + gT_2 - n,\, T_1) + \Theta(p,\, T_2) \big],
\nonumber\\
\Phi_{b,2} &= -\big[ \Theta(p + gT_2,\, T_1) + \Theta(p - n,\, T_2) \big].
\label{eq:branchphaselist}
\end{align}
Equations \eqref{eq:Aout}--\eqref{eq:branchphaselist} display the essential
structure: in each port, the two interfering branches carry identical
Gaussian envelopes---centered at $-gT$ in port $a$ and at
$n - gT$ in port $b$---and differ only by a phase. The composition
\eqref{eq:steps} is what the numerical implementation evaluates directly;
it was verified against an independent split-step Fourier integration of
the same sequence to $2.5\times10^{-13}$ (Appendix~\ref{app:numerics}).

\subsection{Branch phases}
\label{app:branchphases}

The phase differences follow from Eq.~\eqref{eq:Theta} by elementary
algebra. For port $a$,
\begin{align}
\Phi_{a,1} - \Phi_{a,2}
&= \Theta(p + gT_2, T_1) - \Theta(p + gT_2 + n, T_1)
\nonumber\\
&\quad + \Theta(p + n, T_2) - \Theta(p, T_2)
\nonumber\\
&= -\frac{T_1}{2}\big[ n\big( 2(p + gT_2) + n \big) \big]
- \frac{g n T_1^2}{2}
\nonumber\\
&\quad + \frac{T_2}{2}\big[ n(2p + n) \big] + \frac{g n T_2^2}{2}
\nonumber\\
&= -n(T_1 - T_2)\, p - \frac{n g}{2}\big( T^2 - 2T_2^2 \big)
\nonumber\\
&\quad - \frac{n^2 (T_1 - T_2)}{2},
\label{eq:relphasea}
\end{align}
where the last line uses $T_1^2 + 2T_1T_2 - T_2^2 = (T_1+T_2)^2 - 2T_2^2 =
T^2 - 2T_2^2$. The analogous computation for port $b$ gives
\begin{equation}
\Phi_{b,1} - \Phi_{b,2}
= +n(T_1 - T_2)\, p + \frac{n g}{2}\big( T^2 - 2T_2^2 \big)
- \frac{n^2 (T_1 - T_2)}{2}.
\label{eq:relphaseb}
\end{equation}
Restoring dimensions, the coefficient of $p$ is the fringe wave number
$\kappa = \Delta z/\hbar$ with $\Delta z = n\hbar\kO(T_1 - T_2)/m$
[Eq.~\eqref{eq:dz}], the $g$-linear term is the fringe phase $\phi_g$ of
Eq.~\eqref{eq:phase}, and the constant is the recoil phase $\phi_r =
n^2\hbar\kO^2(T_1 - T_2)/2m$. The relative phase is thus linear in
$p$ for all $(n, T_1, T_2, g)$, with no expansion involved, which is the
basis of the fringe form \eqref{eq:fringes}. For later use we also
record the common-mode $g$ derivative of a single branch at $g = 0$,
\begin{equation}
\partial_g \Phi_{a,1}\big|_{g=0}
= -\frac{T^2}{2}\, p - n\Big( \frac{T_1^2}{2} + T_1 T_2 \Big),
\label{eq:commonmode}
\end{equation}
whose $p$-linear part, restored to $-(T^2/2\hbar)\,p$, is the phase imprint
of the $g$-induced \emph{position} displacement $gT^2/2$; it is common to
all four branches and, being a pure phase gradient, is invisible to any
measurement diagonal in momentum (Sec.~\ref{sec:cfi}).

\section{Quantum Fisher information}
\label{app:qfi}

\subsection{Pure-state QFI from the fidelity}
\label{app:qfi-def}

Let $|\psi_g\rangle$ be a normalized pure state depending smoothly on $g$.
Expanding $|\psi_{g+\epsilon}\rangle = |\psi\rangle + \epsilon|\partial_g
\psi\rangle + \tfrac{\epsilon^2}{2}|\partial_g^2\psi\rangle +
O(\epsilon^3)$,
\begin{equation}
\langle \psi | \psi_{g+\epsilon} \rangle
= 1 + \epsilon \langle \psi | \partial_g\psi \rangle
+ \frac{\epsilon^2}{2} \langle \psi | \partial_g^2\psi \rangle
+ O(\epsilon^3).
\end{equation}
Differentiating $\langle\psi|\psi\rangle = 1$ once gives
$\mathrm{Re}\langle\psi|\partial_g\psi\rangle = 0$, and twice gives
$\mathrm{Re}\langle\psi|\partial_g^2\psi\rangle =
-\langle\partial_g\psi|\partial_g\psi\rangle$. Hence
\begin{align}
\big| \langle \psi | \psi_{g+\epsilon} \rangle \big|^2
&= 1 - \epsilon^2 \Big[ \langle\partial_g\psi|\partial_g\psi\rangle
- \big|\langle\psi|\partial_g\psi\rangle\big|^2 \Big] + O(\epsilon^3)
\nonumber\\
&\equiv 1 - \frac{F_Q\,\epsilon^2}{4} + O(\epsilon^3),
\label{eq:fidelity}
\end{align}
which defines the QFI of Eq.~\eqref{eq:qfidef}; the coefficient is fixed by
the standard normalization in which $F_Q = 4\var(\op G)$ for
$|\psi_g\rangle = e^{-ig\op G}|\psi_0\rangle$
\cite{Helstrom1976,Braunstein1994}. Equation~\eqref{eq:fidelity} is also
the basis of the numerical validation of Appendix~\ref{app:numerics}.

\subsection{Generator representation}
\label{app:qfi-gen}

For $|\psi_g\rangle = \op U(g)|\psi_{\rm in}\rangle$ with $|\psi_{\rm
in}\rangle$ independent of $g$, define the Hermitian generator
\begin{equation}
\tilde{\op G} \equiv i\, \op U^\dagger\, \partial_g \op U .
\label{eq:gendef}
\end{equation}
Then $|\partial_g\psi_g\rangle = -i\,\op U \tilde{\op G} |\psi_{\rm
in}\rangle$, so $\langle\partial_g\psi|\partial_g\psi\rangle =
\langle\tilde{\op G}^2\rangle_{\rm in}$ and
$\langle\psi|\partial_g\psi\rangle = -i\langle\tilde{\op G}\rangle_{\rm
in}$, whence
\begin{equation}
F_Q = 4\,\var_{\rm in}\big( \tilde{\op G} \big),
\label{eq:qfivar}
\end{equation}
the variance taken in the \emph{input} state. Note that additive c-numbers
in $\tilde{\op G}$ drop out of the variance.

\subsection{Generator of a free-fall segment}
\label{app:qfi-free}

For $\op U_g(t) = e^{-i\op H t/\hbar}$ with $\op H = \op p_z^2/2m + mg\op
z$, the parameter-differentiation (Duhamel) identity
\begin{equation}
\partial_g e^{-i\op H t/\hbar}
= -\frac{i}{\hbar} \int_0^t e^{-i\op H (t-s)/\hbar}\,
\big( \partial_g \op H \big)\, e^{-i\op H s/\hbar}\, ds
\end{equation}
with $\partial_g\op H = m\op z$ gives
\begin{equation}
\tilde{\op G}_{\rm f}(t) = i\,\op U_g^\dagger \partial_g \op U_g
= \frac{m}{\hbar} \int_0^t \op z_H(s)\, ds,
\end{equation}
where $\op z_H(s) = \op U_g^\dagger(s)\, \op z\, \op U_g(s)$ is the
Heisenberg position.
These follow from $d\op z_H/ds = \op p_{z,H}/m$ and
$d\op p_{z,H}/ds = -mg$: $\op p_{z,H}(s) = \op p_z - mgs$ and $\op z_H(s) =
\op z + \op p_z s/m - g s^2/2$. Performing the elementary integral,
\begin{equation}
\tilde{\op G}_{\rm f}(t)
= \frac{1}{\hbar} \left( m t\, \op z + \frac{t^2}{2}\, \op p_z \right)
- \frac{m g t^3}{6\hbar},
\label{eq:Gfree}
\end{equation}
whose c-number term is dropped henceforth.

\subsection{Pulse conjugations and assembly}
\label{app:qfi-assembly}

Write the full sequence as $\op U = \op K_3 \op U_g(T_2) \op K_2 \op
U_g(T_1) \op K_1$ with $\op K_1 = \op K_3 = \op U_{\pi/2}$ and $\op K_2 =
\op U_\pi$, all $g$ independent. The chain rule applied to
Eq.~\eqref{eq:gendef} gives
\begin{equation}
\tilde{\op G} = \op K_1^\dagger\, \tilde{\op G}_{\rm f}(T_1)\, \op K_1
+ \op K_1^\dagger \op U_g^\dagger(T_1) \op K_2^\dagger\,
\tilde{\op G}_{\rm f}(T_2)\,
\op K_2 \op U_g(T_1) \op K_1 .
\label{eq:chain}
\end{equation}
The required conjugations are computed first. Define the
Hermitian internal operators
\begin{align}
\op\sigma &= |b\rangle\langle b| - |a\rangle\langle a|, \nonumber\\
\op\Sigma_1 &= e^{i n\kO \op z}|b\rangle\langle a|
+ e^{-i n\kO \op z}|a\rangle\langle b|, \nonumber\\
\op\Sigma_2 &= -i\big( e^{i n\kO \op z}|b\rangle\langle a|
- e^{-i n\kO \op z}|a\rangle\langle b| \big),
\end{align}
which satisfy $\op\Sigma_1^2 = \op\Sigma_2^2 = 1$, and note $\op K_2 =
-i\op\Sigma_1$, $\op K_1 = (1 - i\op\Sigma_1)/\sqrt2$. Using $[\op p_z,
e^{\pm in\kO\op z}] = \pm n\hbar\kO\, e^{\pm in\kO\op z}$, direct
computation gives
\begin{align}
\op\Sigma_1 \op p_z \op\Sigma_1 &= \op p_z - n\hbar\kO\,\op\sigma, &
[\op p_z, \op\Sigma_1] &= i\, n\hbar\kO\, \op\Sigma_2,
\nonumber\\
\op\Sigma_1 \op\sigma \op\Sigma_1 &= -\op\sigma, &
i[\op\Sigma_1, \op\sigma] &= 2\,\op\Sigma_2 ,
\label{eq:algebra}
\end{align}
from which the three needed conjugation identities follow:
\begin{align}
\op K_2^\dagger\, \op p_z\, \op K_2
&= \op\Sigma_1 \op p_z \op\Sigma_1 = \op p_z - n\hbar\kO\,\op\sigma,
\label{eq:conjpi}\\
\op K_1^\dagger\, \op p_z\, \op K_1
&= \op p_z + \frac{n\hbar\kO}{2}\big( \op\Sigma_2 - \op\sigma \big),
\label{eq:conjhalfp}\\
\op K_1^\dagger\, \op\sigma\, \op K_1 &= \op\Sigma_2 .
\label{eq:conjhalfsigma}
\end{align}
[Explicitly for Eq.~\eqref{eq:conjhalfp}: $\op K_1^\dagger \op p_z \op K_1
= \tfrac12(1 + i\op\Sigma_1)\op p_z(1 - i\op\Sigma_1) = \tfrac12[\op p_z -
i[\op p_z, \op\Sigma_1] + \op\Sigma_1\op p_z\op\Sigma_1]$, and the algebra
\eqref{eq:algebra} gives the stated result; for
Eq.~\eqref{eq:conjhalfsigma}, $\tfrac12(1+i\op\Sigma_1)\op\sigma(1 -
i\op\Sigma_1) = \tfrac12[\op\sigma + i[\op\Sigma_1,\op\sigma] +
\op\Sigma_1\op\sigma\op\Sigma_1] = \op\Sigma_2$.] The operator $\op z$
commutes with all kicks and is unchanged throughout.

Assembling Eq.~\eqref{eq:chain}: the second term requires first $\op
K_2^\dagger \tilde{\op G}_{\rm f}(T_2)\op K_2 = \tfrac{1}{\hbar}[mT_2\op z
+ \tfrac{T_2^2}{2}(\op p_z - n\hbar\kO\op\sigma)]$ by
Eq.~\eqref{eq:conjpi}; then the free conjugation $\op U_g^\dagger(T_1)
(\cdot) \op U_g(T_1)$, which maps $\op z \to \op z + \op p_z T_1/m -
gT_1^2/2$ and $\op p_z \to \op p_z - mgT_1$ while leaving $\op\sigma$
invariant, giving (c-numbers dropped)
\begin{equation}
\frac{1}{\hbar}\left[ m T_2 \op z + T_2\Big( T_1 + \frac{T_2}{2} \Big)\op
p_z - \frac{n\hbar\kO T_2^2}{2}\,\op\sigma \right];
\end{equation}
and finally the $\op K_1$ conjugation via
Eqs.~\eqref{eq:conjhalfp}--\eqref{eq:conjhalfsigma}. Adding the first term
of Eq.~\eqref{eq:chain}, treated with Eq.~\eqref{eq:conjhalfp} alone, and
collecting coefficients using $T_1^2/2 + T_1T_2 + T_2^2/2 = T^2/2$,
\begin{equation}
\tilde{\op G} = \op G_0(T)
+ \frac{n\kO}{2}\Big( \frac{T^2}{2} - T_2^2 \Big)\, \op\Sigma_2
- \frac{n\kO T^2}{4}\, \op\sigma,
\label{eq:Gtotal}
\end{equation}
with $\op G_0(T) = (mT\op z + \tfrac{T^2}{2}\op p_z)/\hbar$ as quoted below
Eq.~\eqref{eq:QFI}.

The variance of Eq.~\eqref{eq:Gtotal} in the input state $|a\rangle \otimes
|\psi_0\rangle$ is now elementary. Since $\op\sigma|a\rangle =
-|a\rangle$, the $\op\sigma$ term is a c-number on the input state and
contributes nothing. Since $\op\Sigma_2|a\rangle\otimes|\psi_0\rangle =
-i\,|b\rangle \otimes e^{in\kO\op z}|\psi_0\rangle$ is internally
orthogonal to the input, $\langle\op\Sigma_2\rangle = 0$, every cross
covariance $\mathrm{Cov}(\op G_0, \op\Sigma_2)$, $\mathrm{Cov}(\op\sigma,
\op\Sigma_2)$ vanishes (each involves the internal matrix element $\langle
a|\cdots|b\rangle = 0$), and $\var(\op\Sigma_2) = \langle \op\Sigma_2^2
\rangle = 1$. Hence
\begin{equation}
F_Q = 4\var\big(\tilde{\op G}\big)
= 4\var\big( \op G_0 \big)
+ n^2\kO^2 \Big( \frac{T^2}{2} - T_2^2 \Big)^{\!2},
\end{equation}
which is Eq.~\eqref{eq:QFI}. For our Gaussian input,
$\mathrm{Cov}(\op z,\op p_z) = 0$ and $4\var(\op G_0) =
4(mT/\hbar)^2\var(\op z) + (T^2/\hbar)^2 \var(\op p_z)$.

\subsection{Independent check from the branch amplitudes}
\label{app:qfi-branch}

The same result follows directly from the output state,
Eqs.~\eqref{eq:Aout}--\eqref{eq:branchphaselist}, by inserting it into
Eq.~\eqref{eq:qfidef}. Writing (natural units) $\chi_s(p) =
c_s\,\psi_0(p + gT - q_s)\sum_j \varepsilon_{sj} e^{i\Phi_{sj}(p,g)}$ with
$q_a = 0$, $q_b = n$, the $g$ derivative produces an envelope-shift term
$\propto T\psi_0'$ and phase terms $\propto \partial_g\Phi_{sj}$, which are
linear in $p$ [Eq.~\eqref{eq:commonmode} and the relative phases
\eqref{eq:relphasea}--\eqref{eq:relphaseb}]. All required integrals are
then first and second Gaussian moments; cross-branch (oscillatory)
integrals carry the factor $e^{-\Delta z^2\delta p^2/2\hbar^2}$ and are
exponentially negligible. Carrying out the moment algebra---each branch
contributing weight $\tfrac14$, the envelope term $\int(\psi_0')^2 dp =
\sigma^2/2$ per branch, and the phase terms contributing
$\langle(\partial_g\Phi)^2\rangle - \langle\partial_g\Phi\rangle^2$
evaluated over branches and momentum---reproduces
\begin{equation}
F_Q = \frac{T^4}{\hbar^2}\var(\op p_z) + \frac{4m^2T^2}{\hbar^2}\var(\op
z) + \frac{n^2\kO^2}{4}\big( T^2 - 2T_2^2 \big)^2
\end{equation}
(we verified the moment algebra by computer algebra). This route makes
explicit \emph{where} each piece of the QFI resides: the phase term in the
branch-relative phases, the position-variance term in the envelope shift
$mgT$, and the momentum-variance term in the common-mode phase gradient
\eqref{eq:commonmode}. Finally, Eq.~\eqref{eq:QFI} was validated
numerically through the fidelity \eqref{eq:fidelity} of simulated output
states, with agreement at the $10^{-4}$ level across mirrorless,
symmetric, and intermediate configurations for $n$ up to 10
(Appendix~\ref{app:numerics}).

\section{Port distributions and fringe structure}
\label{app:ports}

From Eqs.~\eqref{eq:Aout} and \eqref{eq:Bout},
\begin{align}
P_a(p) &= |A_{\rm out}|^2
= \tfrac14\, \psi_0^2(p + gT)\,
\big| e^{i\Phi_{a,1}} + e^{i\Phi_{a,2}} \big|^2
\nonumber\\
&= \tfrac12\, \psi_0^2(p + gT)\,
\big[ 1 + \cos( \Phi_{a,1} - \Phi_{a,2} ) \big],
\\
P_b(p) &= \tfrac12\, \psi_0^2(p + gT - n)\,
\big[ 1 - \cos( \Phi_{b,1} - \Phi_{b,2} ) \big].
\end{align}
Substituting the relative phases
\eqref{eq:relphasea}--\eqref{eq:relphaseb} and using the evenness of the
cosine yields Eq.~\eqref{eq:fringes} with $Q = \psi_0^2$ (variance $\delta
p^2$), fringe wave number $\kappa = \Delta z/\hbar$, fringe phase $\phi_g$,
and recoil offset $\pm\phi_r$. Three features of these expressions are
used below. (i) The intrinsic contrast is unity: both branches share the
same envelope, so nothing but the detector degrades the fringes. (ii) The
total probability is conserved without invoking any averaging: the
oscillatory parts of $P_a$ and $P_b$ cancel under the momentum integral
only up to exponentially small terms individually, but their sum
$\int(P_a + P_b)\,dp = \int \psi_0^2 = 1$ holds identically because the
port cosines enter with opposite signs and equal envelopes up to the recoil
displacement, whose effect is again exponentially small; the numerics
confirm unit norm to machine precision. (iii) The entire $g$ dependence
resides in the rigid envelope shift $mgT$ and in $\phi_g$; the dispersion
phases $\Theta$ cancel from the distributions except through these two
channels.

\section{Classical Fisher information at finite resolution}
\label{app:cfi}

\subsection{Convolution lemma}
\label{app:conv}

\emph{Lemma.} Let $Q(q) = e^{-q^2/2\delta p^2}/\sqrt{2\pi\delta p^2}$ and
$G_{\sigma_p}(q) = e^{-q^2/2\sigma_p^2}/\sqrt{2\pi\sigma_p^2}$. Then, with
$\Sig^2 = \delta p^2 + \sigma_p^2$,
\begin{align}
\int_{-\infty}^{\infty}\! dq\; G_{\sigma_p}(p - q)\, Q(q)\, e^{i\kappa q}
&= Q_\Sig(p)\; e^{i\kappa p\,\delta p^2/\Sig^2}
\nonumber\\
&\quad\times
e^{-\kappa^2 \sigma_p^2 \delta p^2 / 2\Sig^2},
\label{eq:lemma1}
\end{align}
where $Q_\Sig(p) = e^{-p^2/2\Sig^2}/\sqrt{2\pi\Sig^2}$.

\emph{Proof.} The exponent of the integrand is
\begin{equation}
-\frac{(p-q)^2}{2\sigma_p^2} - \frac{q^2}{2\delta p^2} + i\kappa q
= -A q^2 + B q - \frac{p^2}{2\sigma_p^2},
\end{equation}
with $A = \Sig^2/(2\sigma_p^2\delta p^2)$ and $B = p/\sigma_p^2 + i\kappa$.
The Gaussian integral $\int e^{-Aq^2 + Bq}\,dq = \sqrt{\pi/A}\,
e^{B^2/4A}$ (valid for complex $B$ by contour shift) gives
\begin{equation}
\frac{B^2}{4A} = \frac{\sigma_p^2\delta p^2}{2\Sig^2}
\left( \frac{p}{\sigma_p^2} + i\kappa \right)^{\!2}
= \frac{p^2 \delta p^2}{2\sigma_p^2\Sig^2}
+ \frac{i\kappa p\,\delta p^2}{\Sig^2}
- \frac{\kappa^2\sigma_p^2\delta p^2}{2\Sig^2}.
\end{equation}
Combining the real $p^2$ terms, $\tfrac{p^2\delta
p^2}{2\sigma_p^2\Sig^2} - \tfrac{p^2}{2\sigma_p^2} = -\tfrac{p^2}{2\Sig^2}$,
and collecting the prefactors
$\sqrt{\pi/A}/(2\pi\sigma_p\delta p) = 1/\sqrt{2\pi\Sig^2}$ completes the
proof. $\square$

Setting $\kappa = 0$ recovers the familiar statement that a Gaussian
observed through a Gaussian kernel is a Gaussian of summed variances; the
$\kappa \ne 0$ statement shows that a fringe of wave number $\kappa$
emerges from the blur with (i) contrast multiplied by
$e^{-\kappa^2\sigma_p^2\delta p^2/2\Sig^2}$ and (ii) wave number reduced to
$\kappa_e = \kappa\,\delta p^2/\Sig^2$: the blurred fringe partially locks
to the envelope.

\subsection{Blurred port distributions}
\label{app:blurderive}

Apply the lemma to Eq.~\eqref{eq:fringes}. Writing the fringe as the real
part of $Q(p + mgT)\,e^{i(\kappa p + \phi)}$ with $\phi = \phi_g \pm
\phi_r$, substitute $q' = q + mgT$ in the convolution integral:
\begin{align}
&\int dq\, G_{\sigma_p}(p - q)\, Q(q + mgT)\, e^{i\kappa q}
\nonumber\\
&\qquad = e^{-i\kappa mgT} \int dq'\, G_{\sigma_p}\big( (p + mgT) - q'
\big)\, Q(q')\, e^{i\kappa q'}
\nonumber\\
&\qquad = C_e\; Q_\Sig(p + mgT)\;
e^{i[\kappa_e p \,+\, (\kappa_e - \kappa) mgT]},
\end{align}
using Eq.~\eqref{eq:lemma1} in the last step. Since $\kappa_e - \kappa =
-\kappa\sigma_p^2/\Sig^2$, taking the real part and adding the blurred
envelope term yields Eq.~\eqref{eq:blurred} with the $g$-dependent phase
$\tilde\phi_g$ of Eq.~\eqref{eq:phitilde}: the blurred fringe pattern
acquires an additional $g$ dependence because it partially tracks the
moving envelope. No approximation has been made.

\subsection{Decomposition of the CFI}
\label{app:decomp}

Write the blurred distributions \eqref{eq:blurred} as $\tilde P_\pm =
\tfrac12 \tilde Q\,(1 \pm C_e\cos\tilde\alpha)$ with $\tilde Q(p) =
Q_\Sig(p + mgT)$ and $\tilde\alpha(p) = \kappa_e p + \tilde\phi_g \pm
\phi_r$. Then
\begin{equation}
\partial_g \tilde P_\pm = \tfrac12\Big[ mT\, \tilde Q'\,\big( 1 \pm
C_e\cos\tilde\alpha \big) \mp \tilde Q\, C_e\,
\big(\partial_g\tilde\phi_g\big) \sin\tilde\alpha \Big],
\end{equation}
and therefore
\begin{align}
\frac{\big(\partial_g\tilde P_\pm\big)^2}{\tilde P_\pm}
&= \frac{m^2T^2}{2}\, \frac{\tilde Q'^2}{\tilde Q}\,
\big( 1 \pm C_e\cos\tilde\alpha \big)
\nonumber\\
&\quad \mp\, mT\,\tilde Q'\, C_e \big(\partial_g\tilde\phi_g\big)
\sin\tilde\alpha
\nonumber\\
&\quad +\, \frac{\tilde Q\, C_e^2}{2} \big(\partial_g\tilde\phi_g\big)^2\,
\frac{\sin^2\tilde\alpha}{1 \pm C_e\cos\tilde\alpha} \,.
\label{eq:perport}
\end{align}
Summing the two ports, three simplifications occur, none involving
approximation. (i) The middle (cross) terms carry opposite signs and cancel
pointwise, not merely on average. (ii) The first terms combine to $m^2T^2\,\tilde
Q'^2/\tilde Q$, whose integral is the envelope information: with $\tilde
Q'/\tilde Q = -(p + mgT)/\Sig^2$,
\begin{equation}
m^2T^2 \int \frac{\tilde Q'^2}{\tilde Q}\, dp
= \frac{m^2T^2}{\Sig^4} \int (p + mgT)^2\, \tilde Q\, dp
= \frac{m^2 T^2}{\Sig^2}.
\label{eq:envterm}
\end{equation}
(iii) The last terms combine, using $\tfrac{1}{1+x} + \tfrac{1}{1-x} =
\tfrac{2}{1-x^2}$, into
\begin{equation}
\tilde Q\, \big(\partial_g\tilde\phi_g\big)^2\,
\frac{C_e^2 \sin^2\tilde\alpha}{1 - C_e^2\cos^2\tilde\alpha},
\label{eq:fringeint}
\end{equation}
whose integral is evaluated in the next subsection. [The two ports have
different fringe offsets $\pm\phi_r$ and envelope centers differing by
$n\hbar\kO$; neither affects steps (i)--(iii), which hold pointwise or by
symmetric integration.]

\subsection{Fringe-phase average}
\label{app:avg}

The integral of Eq.~\eqref{eq:fringeint} is of the form $\int \tilde Q(p)\,
W\big(\tilde\alpha(p)\big)\, dp$ with $W$ periodic of period $\pi$. When
the fringe phase winds through many periods across the envelope
(quantified in \ref{app:error}), the integral approaches the uniform average
$\langle W \rangle_\alpha$ times $\int\tilde Q = 1$. The required average
is computed as follows. First, by the tangent half-angle substitution $t =
\tan(\theta/2)$, $d\theta = 2\,dt/(1+t^2)$, $\cos\theta = (1-t^2)/(1+t^2)$,
\begin{align}
\int_0^{2\pi} \frac{d\theta}{A - B\cos\theta}
&= 2\int_{-\infty}^{\infty} \frac{dt}{(A-B) + (A+B)t^2}
\nonumber\\
&= \frac{2\pi}{\sqrt{A^2 - B^2}}, \qquad A > |B| .
\label{eq:weier}
\end{align}
With $1 - k\cos^2\alpha = (1 - \tfrac{k}{2}) - \tfrac{k}{2}\cos 2\alpha$,
Eq.~\eqref{eq:weier} gives $\big\langle (1 - k\cos^2\alpha)^{-1}
\big\rangle_\alpha = (1-k)^{-1/2}$, and with the algebraic identity
\begin{equation}
\frac{\sin^2\alpha}{1 - k\cos^2\alpha}
= \frac{1}{k}\left[ 1 - \frac{1-k}{1 - k\cos^2\alpha} \right]
\end{equation}
one obtains
\begin{equation}
\left\langle \frac{\sin^2\alpha}{1 - k\cos^2\alpha} \right\rangle_\alpha
= \frac{1}{k}\Big[ 1 - \sqrt{1-k}\, \Big]
= \frac{R(k)}{k},
\end{equation}
which is Eq.~\eqref{eq:R} upon $k = C_e^2$. Combining with
Eqs.~\eqref{eq:envterm} and the cancellation of cross terms yields
Eq.~\eqref{eq:master}.

\subsection{Error bound for the fringe average}
\label{app:error}

The sole approximation above is the replacement of $\int \tilde Q\,
W(\tilde\alpha)\,dp$ by $\langle W\rangle_\alpha$. Expanding the periodic
function in its Fourier series, $W(\alpha) = \sum_{l} \hat W_l\,
e^{2il\alpha}$ with $\hat W_0 = \langle W \rangle_\alpha$, the error is
\begin{equation}
\Big| \sum_{l \ne 0} \hat W_l \int \tilde Q(p)\, e^{2il(\kappa_e p +
{\rm const})}\, dp \Big|
\le \sum_{l\ne0} \big|\hat W_l\big|\, e^{-2 l^2 \kappa_e^2 \Sig^2},
\end{equation}
using the Gaussian characteristic function $\big|\int \tilde Q\,
e^{i\lambda p}dp\big| = e^{-\lambda^2\Sig^2/2}$. Since $W(\alpha) =
C_e^2\sin^2\alpha/(1 - C_e^2\cos^2\alpha)$ is analytic in a strip around
the real axis for $C_e < 1$, its Fourier coefficients decay geometrically
and the sum converges to a constant $K(C_e)$ that is finite for $C_e < 1$;
the relative error of Eq.~\eqref{eq:master} is thus bounded by
$K(C_e)\,e^{-2\kappa_e^2\Sig^2}$ at leading harmonic. For the
weakest-contrast entry of Table~\ref{tab:validation} ($n = 1$, $\sigma_p =
4\times10^{-3}$), $2\kappa_e^2\Sig^2 \approx 398$, and for the smallest
fringe number appearing in Fig.~\ref{fig:asym} ($n = 5$, $u = 1/12$),
$2\kappa_e^2\Sig^2 \approx 69$: the exponential factor never exceeds
$e^{-69}$ for any data shown. In the opposite, fringe-poor limit $\Delta z \to 0$
(symmetric KC) no average is needed: the phase is global, $C_e = 1$,
$\kappa_e = 0$, and the per-port expression \eqref{eq:perport} integrates
without approximation to $F_C = (\partial_g\phi_g)^2 + m^2T^2/\Sig^2$ at
any operating
point with $\sin(\phi_g + \phi_r) \ne 0$, since
$\sin^2\!\alpha\,[\tfrac{1}{1+\cos\alpha} + \tfrac{1}{1-\cos\alpha}] = 2$
identically and the cross terms again cancel; this is the KC result used
in Sec.~\ref{sec:optimal}.

\subsection{Leading form and its accuracy}
\label{app:leading}

For $\sigma_p \ll \delta p$, expand Eqs.~\eqref{eq:blurdefs} and
\eqref{eq:phitilde}:
$\Sig^2 = \delta p^2\,[1 + \sigma_p^2/\delta p^2]$;
$C_e^2 = C^2\,\exp[+\kappa^2\sigma_p^4/\Sig^2] = C^2\big[1 +
O(a\,\sigma_p^2/\delta p^2)\big]$ with $a = \kappa^2\sigma_p^2$;
and $\partial_g\tilde\phi_g = \partial_g\phi_g\,\big[1 -
O(\sigma_p^2/\delta p^2)\big]$ for the family (for the mirrorless case
explicitly $\partial_g\tilde\phi_g = \tfrac{n\kO T^2}{2}\,(\delta p^2 -
\sigma_p^2)/\Sig^2$). Substituting into Eq.~\eqref{eq:master} and
discarding all terms of relative order $\sigma_p^2/\delta p^2$ yields the
leading form \eqref{eq:leading}. The deviations of
Eq.~\eqref{eq:leading} from the numerics in Table~\ref{tab:validation}
follow this scaling with an order-unity coefficient, reaching $1\%$ only
for the largest $\sigma_p/\delta p \approx 0.06$ shown, while
Eq.~\eqref{eq:master} agrees with the numerics at the level of their
precision.

\section{Optimization of the momentum transfer}
\label{app:opt}

\subsection{Reduction to a single variable and the fixed point}
\label{app:fixedpoint}

For the mirrorless configuration ($T_2 = 0$, $\Delta z = n\hbar\kO T/m$) in
the design regime $\sigma_p \ll \delta p$, the fringe term of
Eq.~\eqref{eq:leading} is
\begin{equation}
F_{\rm fringe} = \frac{n^2\kO^2 T^4}{4}\, R\big( e^{-a} \big),
\qquad
a = \frac{\sigma_p^2 n^2 \kO^2 T^2}{m^2},
\end{equation}
and eliminating $n^2\kO^2T^2 = a\,m^2/\sigma_p^2$ gives
Eq.~\eqref{eq:hdef} with $h(a) = a[1 - \sqrt{1 - e^{-a}}]$. At the
endpoints, $h(a) = a - a^{3/2} + O(a^2) \to 0$ as $a \to 0$ (using
$\sqrt{1-e^{-a}} = \sqrt{a}\,[1 + O(a)]$) and $h(a) = \tfrac12 a e^{-a}[1
+ O(e^{-a})] \to 0$ as $a \to \infty$; since $h > 0$ in between, an
interior maximum exists. With $w(a) \equiv \sqrt{1 - e^{-a}}$, so that
$w' = e^{-a}/2w$ and $e^{-a} = 1 - w^2$,
\begin{equation}
h'(a) = (1 - w) - \frac{a e^{-a}}{2w}
= (1 - w)\left[ 1 - \frac{a(1+w)}{2w} \right],
\end{equation}
where the factorization uses $a e^{-a}/2w = a(1-w)(1+w)/2w$. Since $w < 1$
for finite $a$, stationarity is equivalent to the fixed-point equation
\eqref{eq:fixedpoint}, $a = 2w/(1+w)$. Numerically (Newton iteration,
converged to the digits quoted),
\begin{equation}
a^* = 0.864036, \quad w^* = 0.760620, \quad h^* = 0.206833,
\end{equation}
with $R(a^*) = 1 - w^* = 0.239380$ and $h''(a^*) = -0.22753 < 0$
confirming a maximum; numerical evaluation of $h'$ shows a single sign
change on $(0,\infty)$, so the maximum is global.

\subsection{Ceiling, floor, and crossover}
\label{app:crossderive}

At the optimum, $F_C(n^*) = (m^2T^2/4\sigma_p^2)\,h^* + m^2T^2/\delta p^2
\simeq 0.05171\, m^2T^2/\sigma_p^2$ in the regime
\eqref{eq:crossover} where the fringe term dominates, which is
Eq.~\eqref{eq:ceiling}. The Cram\'er--Rao bound then gives
\begin{equation}
\Delta g_{\min} = \frac{1}{\sqrt{N F_C(n^*)}}
= \frac{2\sigma_p}{mT\sqrt{N h^*}}
= \frac{4.3976\,\sigma_p}{mT\sqrt{N}},
\end{equation}
Eq.~\eqref{eq:floor}. The fringe strategy beats the envelope information
when $(h^*/4)\,m^2T^2/\sigma_p^2 > m^2T^2/\delta p^2$ (the $\sigma_p^2$ in
$\Sig^2$ being negligible at this order), i.e.\ when $\sigma_p <
\sqrt{h^*}\,\delta p/2 = 0.22739\,\delta p$, Eq.~\eqref{eq:crossover}.

\subsection{Equal-$n$ comparison criterion}
\label{app:critderive}

The mirrorless fringe information exceeds the KC value $n^2\kO^2T_\pi^4$
iff $4R(C^2) > 1$:
\begin{equation}
1 - \sqrt{1 - C^2} > \tfrac14
\;\Leftrightarrow\; 1 - C^2 < \tfrac{9}{16}
\;\Leftrightarrow\; C^2 > \tfrac{7}{16},
\end{equation}
i.e.\ $e^{-a} > 7/16$, or $a < \ln(16/7)$, which is
Eq.~\eqref{eq:criterion} with $\sqrt{\ln(16/7)} = 0.909219$. At the
optimum itself, $4R(a^*) = 4(1 - w^*) = 0.95752 < 1$: since $\sqrt{a^*} =
0.929536$ exceeds the break-even value $0.909219$, the optimized
mirrorless device is marginally inferior to the equal-$n$ KC device, as
discussed in Sec.~\ref{sec:optimal}.

\section{The asymmetric-timing family}
\label{app:asym}

\subsection{Reduction to $f(u)$}
\label{app:freduce}

With $T = 2T_\pi$, $s = uT_\pi$, and $T_2 = T_\pi(1 - u)$,
\begin{equation}
T^2 - 2T_2^2 = 4T_\pi^2 - 2T_\pi^2(1-u)^2 = 2T_\pi^2\big[ 2 - (1-u)^2
\big],
\end{equation}
so the fringe term of Eq.~\eqref{eq:leading} is
\begin{equation}
\frac{n^2\kO^2}{4}\big(T^2 - 2T_2^2\big)^2 R
= n^2\kO^2 T_\pi^4\, \big( 2 - (1-u)^2 \big)^2 R,
\end{equation}
and with $\Delta z = 2n\hbar\kO s/m = 2n\hbar\kO T_\pi u/m$ the contrast
argument is $\sigma_p^2\Delta z^2/\hbar^2 = \beta u^2$ with $\beta$ as in
Eq.~\eqref{eq:family}. Normalizing by the KC value gives $f(u)$.

\subsection{Stationarity condition}
\label{app:ustat}

With $x(u) = e^{-\beta u^2}$, $dR/dx = 1/(2\sqrt{1-x})$ and $dx/du =
-2\beta u x$, so $dR/du = -\beta u x/\sqrt{1 - x}$ and $f'(u) = 0$ reads
\begin{equation}
4(1 - u)\big( 2 - (1-u)^2 \big)\, R(x)
= \big( 2 - (1-u)^2 \big)^2\, \frac{\beta u\, x}{\sqrt{1 - x}},
\label{eq:ustatcond}
\end{equation}
a transcendental equation solved numerically for the optima shown in
Fig.~\ref{fig:asym}.

\subsection{Nonanalytic expansion and the near-threshold optimum}
\label{app:uexp}

For small $u$ at fixed $\beta$, expand both factors. The prefactor is a
polynomial, $\big(2 - (1-u)^2\big)^2 = (1 + 2u - u^2)^2 = 1 + 4u + 2u^2 -
4u^3 + u^4$. For the contrast factor, $1 - x = \beta u^2 - \tfrac12
\beta^2 u^4 + O(\beta^3u^6)$, so
\begin{equation}
\sqrt{1 - x} = \sqrt{\beta}\,|u|\Big[ 1 - \frac{\beta u^2}{4} +
O(\beta^2u^4) \Big],
\end{equation}
and $R = 1 - \sqrt{\beta}\,|u| + \tfrac14\beta^{3/2}|u|^3 +
O(\beta^2u^4)$: the square root of $1 - x$ is nonanalytic in $u$ at $u =
0$, which is inherited by $f$. Multiplying the two expansions for $u > 0$
gives Eq.~\eqref{eq:fexp}. The one-sided derivative $f'(0^+) = 4 -
\sqrt{\beta}$ yields the threshold criterion \eqref{eq:asymcriterion}.
Near threshold, writing $\varepsilon = 4 - \sqrt{\beta}$ with $0 <
\varepsilon \ll 4$ and truncating Eq.~\eqref{eq:fexp} at quadratic order,
\begin{equation}
f'(u) = \varepsilon - 2\big( 4\sqrt{\beta} - 2 \big) u + O(u^2) = 0
\;\;\Rightarrow\;\;
u^* = \frac{\varepsilon}{8\sqrt{\beta} - 4},
\end{equation}
which is Eq.~\eqref{eq:ustar}; the truncation is consistent because $u^*
= O(\varepsilon)$ and $\beta u^{*2} = O(\varepsilon^2) \ll 1$ there. For
all interior optima shown in Fig.~\ref{fig:asym}, the separation satisfies
$\sigma_p\Delta z^*/\hbar = \sqrt{\beta}\,u^* \lesssim 0.4$: the optimum
holds the wave-packet separation safely below the detector's resolving
length, trading a controlled fraction of the phase enhancement for fringe
visibility.

\section{Numerical methods and validation protocols}
\label{app:numerics}

\emph{Direct composition.} All numerics implement the closed-form
composition of Appendix~\ref{app:dynamics}\,\ref{app:composition}
directly: the output amplitudes are evaluated on a momentum grid of
spacing $10^{-4}\,\hbar\kO$ (at least ten points per fringe for all $n \le
30$, refined to $5\times10^{-5}\,\hbar\kO$ for the $n > 30$ points of
Fig.~\ref{fig:optimal}(b)) with no time discretization. The composition
was verified against an independent split-step Fourier integration of the
same sequence in position space: the output-port populations agree to
$2.5\times10^{-13}$ across two full fringes of $g$.

\emph{Detection and CFI.} Detection resolution is applied by
Fourier-domain Gaussian convolution of the port distributions, and the
CFI \eqref{eq:cfidef} is computed by central finite differences in $g$
with a step chosen so that the fringe phase changes by $10^{-4}$~rad; this
step sets the $\sim10^{-8}$ relative floor quoted in
Table~\ref{tab:validation}. Mirrorless and asymmetric configurations are
evaluated at $g \to 0$; for the symmetric KC endpoint, where the fringe
phase is global, the CFI is evaluated at the mid-fringe operating point
$n\kO g_0 T_\pi^2 = \pi/2$, as in experimental practice, thereby avoiding
the vanishing first derivative at the fringe extremum. (Only in this
$\Delta z = 0$ limit does the operating point---equivalently a laser
phase---matter; for $\Delta z \ne 0$ the uniform fringe-phase sampling
renders the CFI independent of it, as Eq.~\eqref{eq:master} makes
explicit.)

\emph{QFI validation.} The QFI \eqref{eq:QFI} was checked independently of
any measurement model through the fidelity \eqref{eq:fidelity} between
simulated output states at $g$ and $g + \epsilon$, with $\epsilon =
1.5\times10^{-7}$ chosen so that $1 - |\langle\psi(g)|\psi(g +
\epsilon)\rangle| \sim 10^{-6}$ sits far above the double-precision floor:
the extracted $F_Q$ agrees with Eq.~\eqref{eq:QFI} to better than
$10^{-4}$ (relative) for the mirrorless, symmetric, and
intermediate-asymmetry configurations at $n$ up to 10.

\emph{Parameter degeneracy.} One caution on parameter choice: the default
$(\sigma\kO, T/t_0) = (10, 200)$ accidentally satisfies $T\,\delta
p^2/(m\hbar) = 1$, which makes the two motional contributions
$(T^2/\hbar)^2\var(\op p_z)$ and $4(mT/\hbar)^2\var(\op z)$ numerically
equal and would mask a misidentification of the envelope information. The
$\sigma_p \to 0$ and finite-$\sigma_p$ validations were therefore repeated
at $(\sigma\kO, T/t_0) = (10, 100)$, $(10, 400)$, and $(5, 200)$, where
the two pieces differ by factors of $4$--$16$: in all cases the envelope
information equals $m^2T^2/\delta p^2$ (the position-variance term, as
stated in Sec.~\ref{sec:cfi}), and the closed forms remain accurate at the
stated levels. The initial state is a Gaussian of width $\sigma =
10\,\kO^{-1}$ and the total time is $T = 200\,t_0$ throughout the figures,
matching Ref.~\cite{Kritsotakis2018}.

\end{document}